\documentclass[12pt,a4paper,twocolumn]{narms2}

\usepackage{subfigure} 
\usepackage{psfig}
\usepackage{timesmt}   
\usepackage{chikako}
\usepackage{graphicx}
\usepackage{epsfig}
\usepackage{bm}

\usepackage{amssymb}
\makeatletter
\renewcommand{\section}{\@startsection%
{section}%
{1}%
{0mm}%
{- \baselineskip}%
{0.15\baselineskip}%
{\normalfont\normalsize}}%

\renewcommand{\subsection}{\@startsection
{subsection}%
{2}%
{0mm}%
{-\baselineskip}%
{0.15\baselineskip}%
{\normalfont\normalsize}}%
\makeatother

\begin{document}

\title{The hierarchical properties of contact networks in granular packings }

\author{ T. Aste and T.J. Senden 
\em Department of Applied Mathematics, 
\\ Research School of Physical Sciences and Engineering, 
\\ The Australian National University, 0200 Australia.
}

\vspace{-3.cm}

\abstract{
\vspace{-2.5cm}
{\it proceedings of Powders and Grains 2005}
\\
\\
The topological structure resulting from the network of contacts between grains (\emph{contact network}) is studied for very large samples of disorderly-packed monosized spheres with densities ranging from 0.58 to 0.64.
The hierarchical organization of such a structure is studied by means of a topological map which starts from a given sphere and moves outwards  in concentric shells through the contact network.
We find that the topological density of \emph{disordered} sphere packing is larger than the topological density of equivalent \emph{lattice} sphere packings. 
}



\maketitle
\frenchspacing

\section{\sc Introduction}
\label{s.I}

To identify the accessible configurations at the grain level  and to understand  which are the possible combinations of such local configurations which generate the global packing is  of singular importance, hence it is the necessary starting point for any fundamental understanding of granular matter.
Until now the empirical investigation of the geometrical structure of granular packings have been limited by the paucity of accurate experimental data.
Indeed, after the seminal works of Bernal, Mason and Scott \shortcite{Bernal60,Scott62}, it has been only very recently that the use of tomography has allowed one to `see' the three dimensional structure of such systems and explore their geometry from the grain level up to the whole packing \shortcite{Seidler00,Sederman01,Richard03,AsteKioloa,Aste04}.

In this paper we report and discuss results from an empirical investigation by means of X-ray Computed Tomography on very large samples of disorderly packed monosized spheres with densities\footnote{The density is defined as the fraction of the volume occupied by  the balls divided by the total volume of the region of the space considered.}  ranging from 0.58 to 0.64. 
This study is the largest and the most accurate empirical analysis of disordered packings at the grain-scale ever performed.
A detailed description of the experimental methodology and apparatus are reported in \shortcite{AsteKioloa,Aste04}.
We analyzed 6 samples of monosized acrylic spheres  in cylindrical containers with roughened walls,  and with an inner diameter of $ 55\; mm$, filled to a height of $\sim 75\; mm$. 
In order to verify possible effects due to gravity and sizes we used two kind of acrylic spheres with diameters $d =1.00 \; mm $ and  and $d = 1.59 \; mm $, with polydispersities within $0.05 \; mm$.
The smaller spheres were used to prepare two samples (referred as sample A and C hereafter) at densities 0.586 and 0.619 containing about 103,000 and 143,000 spheres,  respectively. 
Whereas,  the larger spheres were used to prepare four samples at densities 0.596, 0.626 0.630 and 0.64 (samples B, D, E and F) each containing about 35,000 beads.
A X-ray Computed Tomography apparatus (see Sakellariou \textit{et al}.  \shortcite{Sakellariou04}) was used to image the samples.  
The two large samples (A, C) were analysed by acquiring data sets of $2000^3$ voxels with a spatial resolution $0.03 \;mm$; whereas for the other four samples (B, D, E and F) were acquired data sets of $1000^3$ voxels with a spatial resolution $0.06 \; mm$.
The positions of the center for each sphere were retrieved,  with a sub-voxel precision,  by a convolution method applied to the segmented  \shortcite{Sheppard04} datasets (see \shortcite{AsteKioloa,Aste04} for further details).

\begin{figure}
\begin{center}
\resizebox{0.40\textwidth}{!}{%
\includegraphics{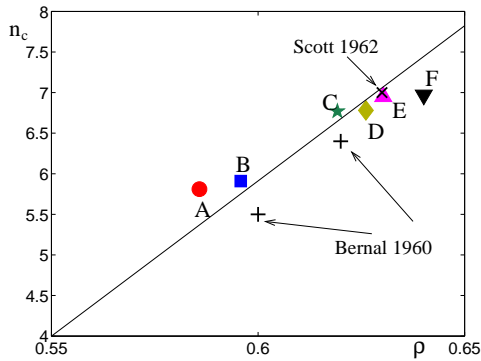} 
}
\end{center}
\caption{ 
\label{f.Nc_rho}
 \small
Average coordination number vs. sample density.
The filled symbols correspond to the samples investigated in the present work.
The two symbols `$+$' are the values from Bernal 1960; whereas the `$\times$' is from Scott 1962.
The line is the best linear  fitting of the A-F data constrained to  pass through $n_c = 4$ at $\rho = 0.55$.
}
\end{figure}

\section{\sc Coordination Number}
\label{s.NNC}\label{s.5}

\begin{figure}
\begin{center}
\resizebox{0.40\textwidth}{!}{%
\includegraphics{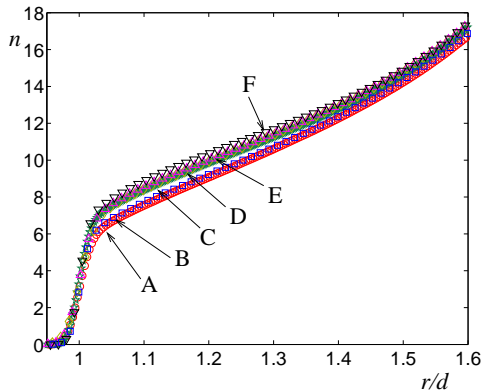} 
}
\end{center}
\caption{ 
\label{f.c_cum}
 \small
The average coordination number increases with the radial distance $r$ within which two spheres are considered in contact.
 }
\end{figure}

The average number of spheres in contact with any given sphere have been widely investigated in the literature of granular matter \shortcite{Bernal60,Scott62,Seidler00,ppp,Sederman01,Silb02,Donev04c}.
Indeed, this is a very simple topological quantity which gives  important information about the local configurations and the packing stability  and determines the cohesion of the material.
The average number of spheres within a radial distance $r$ from a given sphere ($n_t(r)$) increases with the radial distance.
Its empirical behavior is shown in Fig.\ref{f.c_cum}.
One can note that above $r=0.98 d$ the number of neighbors grows very steeply up to a `knee' at about $1.02 d$  where a slower growth takes place.
The definition of the coordination number depends therefore on the radial distance within which two spheres are considered in contact.
However, a deconvolution method developed in \shortcite{Aste04} can be used to distinguish between the contribution from the spheres in contact and the contribution from other spheres which are  near but do not touch.
We estimate that in the 6 samples A-F the average number of spheres in contact  $n_c$ are between 5.81 and 6.97.
In fig.\ref{f.Nc_rho} the values of $n_c$ are  reported as a function of the sample densities showing a clear and consistent  increasing trend with the density. 
Such dependence on the packing density (Fig.\ref{f.Nc_rho}) has important conceptual implications which are discussed in \shortcite{Aste04}.

In this paper we construct `contact' networks by using different thresholds on the distance within which two spheres are considered in contact. This produces several different contact networks with average coordinations ($n$) that increase with the threshold distance.

\section{\sc  Topological Structure Beyond First Neighbors}
\label{s.CNBFN}\label{s.7}

Any force-path or any infinitesimal local grain displacement must mechanically propagate from grain to grain through the network of touching grains.
The understanding of the hierarchical organization of such contact network beyond first neighbors is therefore crucial.
Here we apply to granular matter an approach which was originally developed for the study of crystalline systems \shortcite{Brunner71,OKeeffe91,Conway97} and disordered foams \shortcite{ASBORI}.
The topological structure of crystalline frameworks has been intensely studied in terms of the number of atoms that are $j$ bonds away from a given atom.
If we start from a given `central' atom, the first `shell' (distance $j=1$) is made by all the atoms in contact with the central one.
The second shell (distance $j=2$) consists of all atoms which are neighbors to the atoms in the first shell, excluding the central one.
Moving outward, the  atoms at shell $j+1$  are all the ones which are bonded to atoms in shell $j$ and which have not been counted previously.
In infinite, periodic, crystalline structures with no boundaries,  the number of atoms per shell should increase with the topological distance and it has been shown that in several three-dimensional crystalline structures  the law of growth for the number of atoms ($K_j$) at shell $j$ can be described with: $K_j = a_j j^2 + b_j j + c_j$  (with $a_j$,  $b_j$ and $c_j$ coefficients that might vary with $j$ but only within a bounded finite interval) \shortcite{Brunner71,OKeeffe91,Conway97,Grosse96}.
Following the definition of O'Keeffe \shortcite{OKeeffe91b}, for these crystalline systems, the asymptotic behavior  of $K_j$ can be characterized in terms of an `exact topological density' : $TD = \left< a_j \right>/3$ \shortcite{Grosse96}.
It has been noted that such a  topological density is interestingly related to the geometrical density of the corresponding crystalline structure and it is a powerful instrument to characterize such systems.
For instance, it is easy to compute that the cubic lattice has $K_j = 4 j^2 +2$.
Whereas, spheres packed in a $bcc$ (body centered cubic) crystalline arrangement have: $K_j = 6 j^2 +2$ ($j>0$). 
On the other hand, it has been shown \shortcite{Conway97} that for Barlow packings of spheres,  $K_j$ are always in a narrow range within:
\begin{equation}
\label{barlow}
10 j^2 + 2 \le K_j \le \lfloor \frac{21 j^2 }{2} \rfloor + 2 \;\;\;\; (j > 0)\;\;,
\end{equation}
where the brackets $ \lfloor ...  \rfloor$ indicate the floor function.
It has been observed by O'Keeffe and Hyde \shortcite{OKeffeBook} that for  \emph{lattice sphere packings} with coordination number $n$, the general rule holds:  $K_j = (n -2) j^2  + 2$, implying therefore $a = n -2$.  

\begin{figure}
\begin{center}
\resizebox{0.40\textwidth}{!}{%
\includegraphics{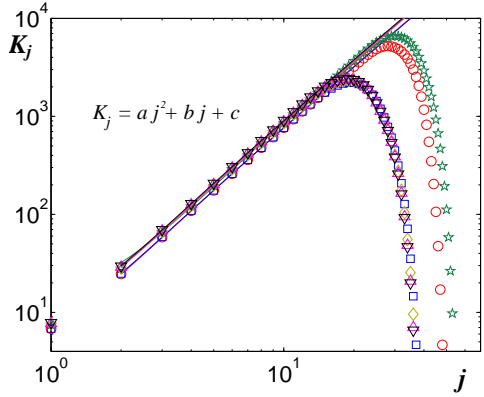} 
}
\end{center}
\caption{ \small
Shell occupation numbers vs. topological distance. 
The symbols indicate the different samples (as in Fig.\ref{f.Nc_rho}) and the lines are the best-fits (of the growing part only) using the polynomial form: $K_j = a  j^2 + c_1 j + c_0$.
The fits are between $j=2$ and $\hat j =$ 10 (for samples B, D, E, F ) and $\hat j =$ 15 (for samples A, C).
The data refer to threshold distance  $1.05d$. 
 }
\label{f.shell}
\end{figure}

Beyond perfect crystalline order very few results are known either from theoretical, empirical or numerical point of view.
One can argue that $K_j$ must grow with a law comparable with the law for a spherical shell: $K_j \sim a j^2 \sim 4 \pi j^2$.
However, it is also clear that the shape of the growing shell and its roughness can drastically change the coefficient $a$ (as observed in two dimensional cases \shortcite{asteSoap96}).
Moreover, it can be shown \shortcite{ADHhypnet04} that in some topological networks the law of growth can follow an intrinsic dimension which is different from the dimension of the embedding space (3 in our case).
This mechanism can produce power law growth with exponents different from 2, or different behaviors such as exponential -- or even faster --  laws of growth  \shortcite{ADHhypnet04}. 

We observed that the number of spheres at a given topological distance $j$ from a central one follows a power law growth (see Fig.\ref{f.shell}) until a critical distance $\hat j$, above which the shells hit the sample boundaries and $K_j$ starts to decrease.  
We verify that a quadratic law $K_j = a j^2 + c_1 j + c_0$ fits quite accurately  the observed behaviors of $K_j$ for $j < \hat j$.
This fixes the intrinsic dimension for these systems equal to 3 (which coincides with the geometrical dimension of the embedding space).

\begin{figure}
\begin{center}
\resizebox{0.40\textwidth}{!}{%
\includegraphics{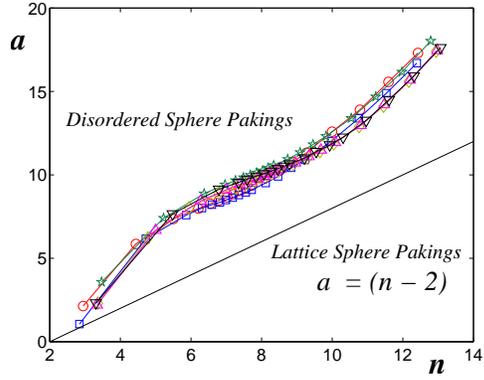} 
}
\end{center}
\caption{
 \small
 The coefficient $a$ plotted against the average coordination number of the contact network ($n$) shows that disordered packings have larger topological densities in comparison with \emph{lattice} sphere packings. 
}
\label{f.shell2}
\end{figure}

We find that the coefficient $a$ depends on the threshold distance within which neighboring spheres are considered in contact.
Indeed, changes in the threshold distances are unavoidably associated with changes in the contact network and an enlargement of the threshold distance must correspond to a thickening of the shell. 
It has been noted previously that such a threshold affects also the average coordination number $n$ in the contact network (Fig.\ref{f.c_cum}).
In Fig. \ref{f.shell2} we show that these two quantities are positively correlated: the coefficient $a$ increases monotonically with $n$.
Interestingly, a comparison with the known behavior in lattice sphere packings ($a = n-2$) shows that these disordered systems are topologically more compact than the analogous lattices.
This observation might be relevant when the structural stability and rigidity of such system are concerned.
Understanding whether this is a bare consequence of topological disorder or it is associated to the properties of such granular systems is a challenging topic which must be explored.

\section{\sc Conclusion}

The study of the topological structure of the contact network at the grain level shows that the average number of spheres in contact increases almost linearly with packing density and  lies between 5.5 and 7.5 in the range of densities between $0.58$ and $0.64$.
An extrapolation to the random loose packing density ($\rho = 0.55$) suggests that at this density the system could have an average number of 4 neighbors per sphere.
This implies the possibility of a rigidity percolation transition taking place at the random loose packing limit.

The structure beyond first neighbors, studied by means of a topological map, shows  that the contact network has an intrinsic dimension of 3.
Surprisingly, we found that the topological density in disordered sphere packings is always larger than the topological density in the corresponding lattice sphere packings.
Such a larger topological density is an indication that the contact network is more compact in disordered systems despite the fact that the \emph{geometrical}  density is lower.
This fact might have important implication on the system stability and resilience to perturbations and shocks.
Notably, this fact is consistent with  what was observed in \shortcite{Aste04} where we found that in a certain region around the grains, disordered packings can be locally more compact than the crystalline ones.

\subsection*{Acknowledgements}
Many thanks to M. Saadatfar, A. Sakellariou for the tomographic data and several discussions.
This work was partially supported by the ARC discovery project DP0450292 and Australian Partnership for Advanced Computing National Facilities (APAC).


\bibliographystyle{chikako}

\end{document}